\documentclass[sigconf, nonacm]{acmart}
\usepackage[normalem]{ulem}

\usepackage{amsmath,amsfonts}
\usepackage{algorithmic}
\usepackage{graphicx}
\usepackage{textcomp}
\usepackage{xcolor}
\usepackage{epsfig}

\usepackage{enumerate}

\usepackage{graphicx}
\usepackage{balance}
\usepackage{comment}
\usepackage{listings}
\usepackage{color,soul}
\usepackage{colortbl}
\usepackage{moreverb}
\usepackage{framed}
\usepackage{multirow}
\usepackage{pgfplots}
\usepackage{multirow}
\usepackage{rotating}
\usepackage{makecell}
\usepackage{pifont}
\usepackage{array}
\usepackage{subfig}

\usepackage{tcolorbox}

\usepackage{tikz}

\makeatletter
\newif\if@restonecol
\makeatother

\usepackage[ruled,vlined,linesnumbered]{algorithm2e}
\definecolor{lightgray}{gray}{0.9}
\definecolor{lightblue}{rgb}{0.9,0.9,1}
\definecolor{red}{rgb}{1,0,0}

\newcommand\cut[1]{}

\newcolumntype{L}[1]{>{\raggedright\let\newline\\\arraybackslash\hspace{0pt}}m{#1}}

\newcommand{\crossmark}{\ding{55}}
\newcommand{\chkmark}{\ding{51}}

\newcommand\os{operating system\xspace}
\newcommand\oses{operating systems\xspace}

\newcommand\Sname{Casper\xspace}

\begin{document}

\title{\Sname: Prompt Sanitization for Protecting User Privacy in Web-Based Large Language Models}

\author{Chun Jie Chong}
\affiliation{
\institution{\large New Jersey Institute of Technology}
\city{Newark}
\state{New Jersey}
\country{USA}
}
\email{cc255@njit.edu}

\author{Chenxi Hou}
\affiliation{
\institution{\large New Jersey Institute of Technology}
\city{Newark}
\state{New Jersey}
\country{USA}
}
\email{ch395@njit.edu}

\author{Zhihao (Zephyr) Yao}
\affiliation{
\institution{\large New Jersey Institute of Technology}
\city{Newark}
\state{New Jersey}
\country{USA}
}
\email{zhihao.yao@njit.edu}

\author{Seyed Mohammadjavad Seyed Talebi}
\affiliation{
\institution{\large Pyte}
\city{Los Angeles}
\state{California}
\country{USA}
}
\email{mjavad@uci.edu}


\begin{abstract}

Web-based Large Language Model (LLM) services have been widely adopted and have become an integral part of our Internet experience.
Third-party plugins enhance the functionalities of LLM by enabling access to real-world data and services.
However, the privacy consequences associated with these services and their third-party plugins
are not well understood.
Sensitive prompt data
are stored, processed, and shared by cloud-based LLM providers and third-party plugins.
In this paper, we propose \Sname, a prompt sanitization technique that aims to protect user privacy by detecting and removing sensitive information from user inputs before sending them to LLM services.
\Sname runs entirely on the user's device as a browser extension and does not require any changes to the online LLM services.
At the core of \Sname is a three-layered sanitization mechanism consisting of a rule-based filter, a Machine Learning (ML)-based named entity recognizer, and a browser-based local LLM topic identifier.
We evaluate \Sname on a dataset of 4000 synthesized prompts and show that it can effectively filter out Personal Identifiable Information (PII) and privacy-sensitive topics with high accuracy, at 98.5\% and 89.9\%, respectively.
\end{abstract}

\settopmatter{printacmref=false}
\settopmatter{printfolios=true}
\pagenumbering{arabic}

\maketitle

\section{Introduction}
\label{sec:intro}

Large Language Model (LLM) has been widely adopted in various online applications, such as chatbots, search engines, and translation tools.
The success of LLM is a result of recent advancements in machine learning, allowing transformer models~\cite{vaswani2017attention} to train on a large corpus of data, measured in tens and hundreds of trillion tokens~\cite{touvron2023llama}.
Cloud-based training and inference services have made LLM readily accessible to regular users, who can utilize the powerful language models without maintaining prohibitively expensive computational resources.

Cloud-based LLM services offer convenience, but unfortunately, they also raise privacy concerns.
As they are black-box systems, users have little control over how their prompts are processed and stored.
According to their privacy policies,
LLM service providers
collect the prompts and share user's inputs with third-party service providers
for business-related purposes
~\cite{microsoft_privacy_statement, openai_privacy_policy}.
We will take a closer look at the privacy policies of mainstream LLM service providers
in Section~\ref{sec:bg_llm}.

Indeed, the prompt collection is essential for improving the quality of LLM responses.
Training data for LLM models are collected from a broad range of sources to ensure the model's generalization and flexibility.
Without proper context, LLM may generate irrelevant or even nonsensical responses~\cite{xu2024hallucination}.
Using
previous prompts to fine-tune and perform in-context learning improves
the coherence and relevance of subsequent responses,
effectively
serving as a few-shot learning mechanism~\cite{touvron2023llama},
but these features raise privacy concerns, as the refined model may leak sensitive information
~\cite{duan2023privacy}.

In addition, LLM plugins, such as online shopping portal and travel arrangement service, provide rich functionalities that allow LLM to interact with the real world, and these plugins may require users to share their inputs with third-party service providers to fulfill the requested tasks~\cite{gpt_plugins, copilot_plugins}.
Once the user's inputs are shared with third-party service providers, the providers may store and
process them in accordance with their respective privacy policies.
For example,
Expedia has developed a plugin for ChatGPT for streamlined travel booking through LLM chat experience~\cite{expedia_plugin}.
When users enable the plugin and ask for travel accommodation, their prompts are processed by Expedia.
Although Expedia's privacy policy does not specifically mention LLM prompts, it does state data that enables booking may be collected and used~\cite{expedia_privacy}.

\begin{figure*}
\centering
\subfloat[Online LLM services.] {
\centering
\includegraphics[height=0.23\textheight]{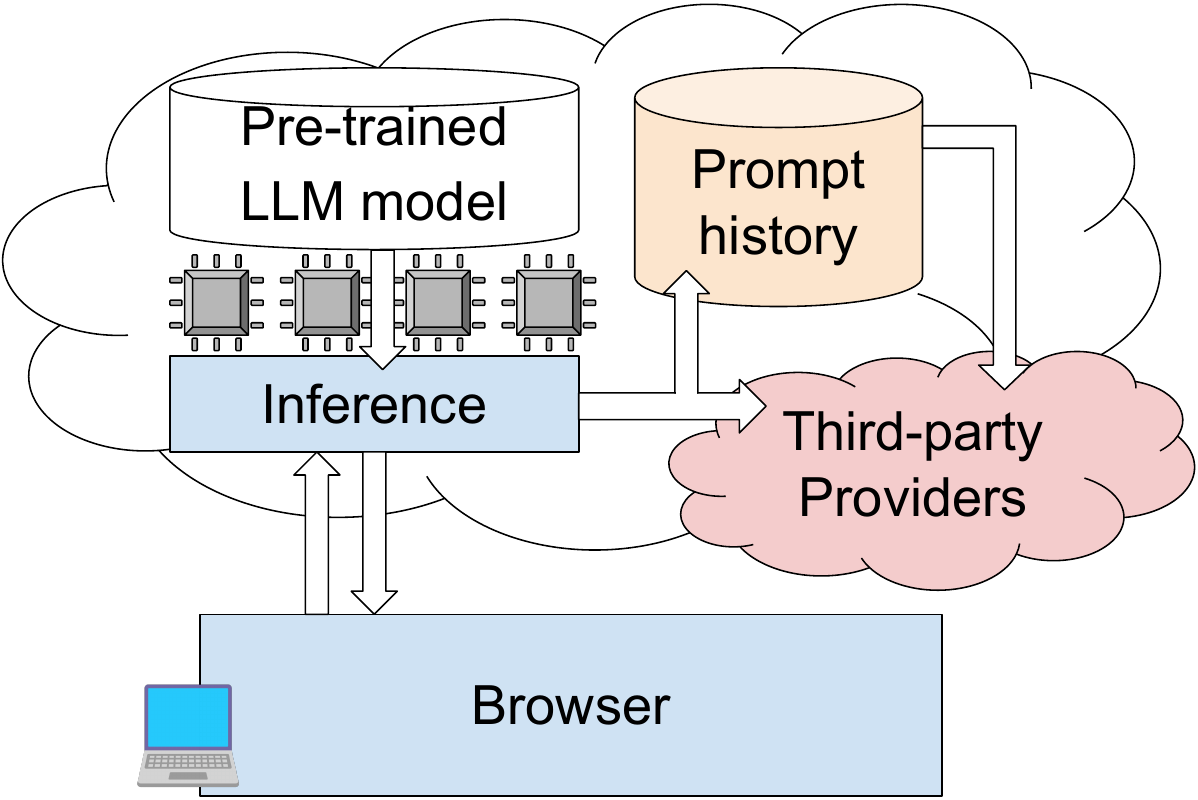}
\label{fig:llm_services}
}
\hfill
\subfloat[\Sname architecture.] {
\includegraphics[height=0.23\textheight]{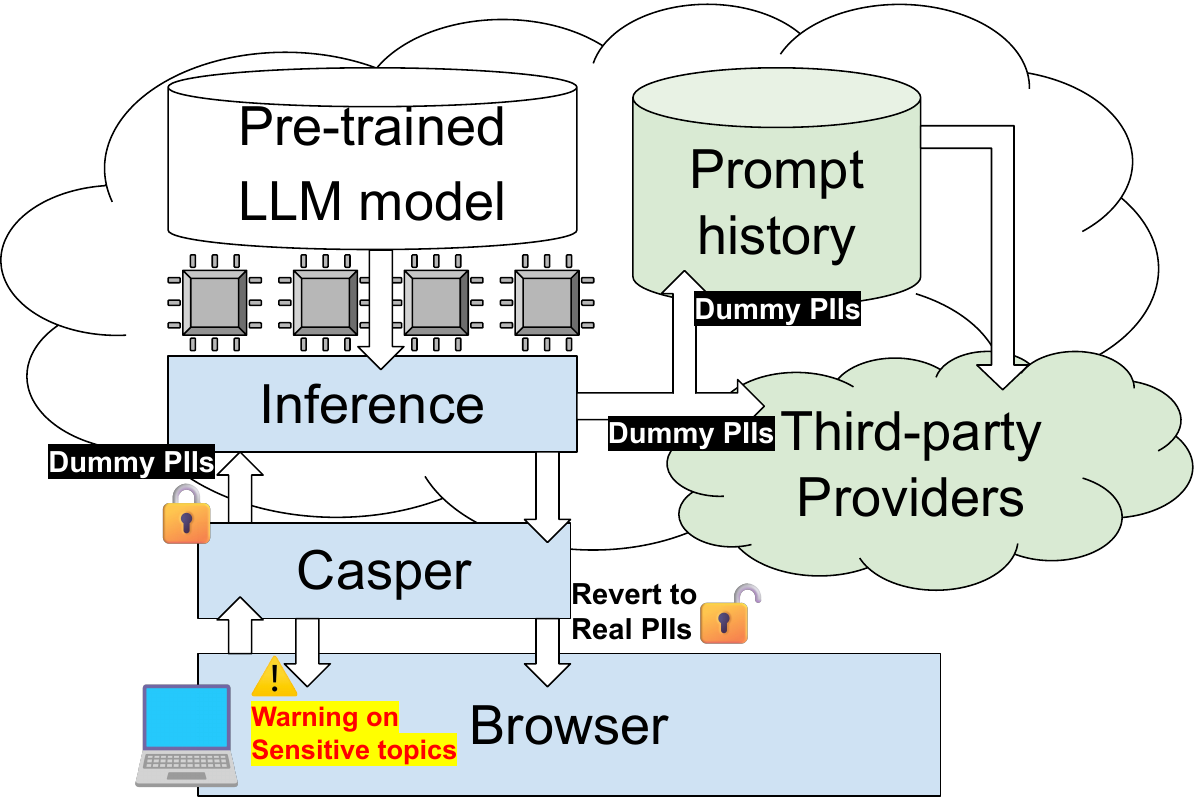}
\label{fig:llm_ours}
}

\caption{Comparison of online LLM services and \Sname architecture.
Blue represents trusted components, green represents minimal privacy risk, yellow represents increased privacy risk, and red represents substantial privacy risk.}
\label{fig:llm_all}
\end{figure*}

Similar privacy concerns have been raised and addressed in voice assistant systems~\cite{seyedtalebi2021_2}.
Voice assistants constantly listen to users' voice commands, and send them to the cloud for processing.
If requests are intended for third-party services, their transcript may be sent to third-party
skills for further processing~\cite{alexa_skill}.
Megamind~\cite{seyedtalebi2021_2} addresses this privacy issue by interposing on the voice commands and filtering out sensitive information locally, so that only non-sensitive information is shared with the cloud and third-party skills.

To address the urgent need for privacy-preserving online LLM services,
we design and implement a lightweight, yet effective,
privacy-preserving system for online LLM services, called \Sname.
\Sname is a browser extension
that runs entirely on user's device.
The sanitization layers in \Sname filter out sensitive information in user's prompts before they are sent out to LLM service providers.
\Sname does not require any modification to LLM services.
Our prototype is compatible with OpenAI ChatGPT's web interface, and can be easily extended to other online LLM services.

At the core of \Sname is a three-layered sanitization mechanism consisting of a rule-based filter, a Machine Learning (ML)-based named entity recognizer, and a local WebGPU-based LLM topic identifier.
All three layers of filtering are designed to be lightweight and efficient, so that \Sname runs in a browser extension without using any cloud-based services or requiring any software installation.

First,
a rule-based filter identifies and redacts sensitive information based on pre-defined keywords and generic matching rules for common sensitive information.
These rules can be customized by users, for example, to include their names, email addresses, and customized matching rules.
Second,
a machine learning (ML)-based detector identifies named entities, such as names, locations, and organizations.
Finally, to further enhance privacy protection, a WebGPU-based local LLM identifies privacy-sensitive topics in the prompts and warns users about potential privacy risks.
User's prompts may leak privacy-sensitive information without containing any Personal Identifiable Information (PII) or sensitive keywords.
For example, a user may ask medical questions about their pregnancy without mentioning a specific medical terminology: ``I feel dizzy and nauseous in the morning at the 10th week.''
Rule-based and ML-based filters may not be able to detect such privacy-sensitive topics, but the topic identifier can.

We evaluate \Sname on a set of 4000 synthesized prompts.
These synthesized prompts cover a wide range of information, from privacy-sensitive prompts, including names, locations, organizations, medical information, and legal information,
to non-privacy-related prompts, such as generic questions and prompts without named entities.
Our evaluation shows that \Sname is highly effective in identifying PII in user's prompts at 98.5\% accuracy.
Furthermore, \Sname only incurs a small performance overhead in using the browser-based LLM inference in detecting privacy-sensitive topics at 89.9\% accuracy.

We summarize the research contributions of \Sname as follows:
\begin{itemize}
\item We study the privacy issues in online LLM services and their third-party plugins.
\item We design and implement a lightweight browser extension-based system for preserving user privacy in using online LLM services. To our knowledge, \Sname is the first client-side system addressing privacy issues in the LLM web interface.
\item We show that rule-based, ML-based, and LLM-based prompt sanitization mechanisms effectively complement each other in identifying privacy-sensitive information in user's prompts.
\item We demonstrate \Sname is highly efficient and effective in protecting user privacy, with minimal impact on user experience and zero modification to LLM web interfaces.
\end{itemize}

The rest of this paper is structured as follows, Section~\ref{sec:bg} introduces the background of online LLM services and their privacy concerns.
Section~\ref{sec:model} describes our threat model and trust assumptions.
Section~\ref{sec:motivation} discusses the motivation and design goals of \Sname.
Section~\ref{sec:design} presents the architecture of our three-layered filtering mechanism and other research challenges.
Section~\ref{sec:impl} describes the implementation details of \Sname as a Chrome browser extension.
Section~\ref{sec:eval} evaluates \Sname's effectiveness and performance overhead.
Section~\ref{sec:related} draws comparisons with related work, and Section~\ref{sec:discussion} discusses limitations and future work.
Finally, Section~\ref{sec:conclusion} concludes the paper.
\section{Background}
\label{sec:bg}

\begin{table*}[]
\begin{tabular}{l|lllllllllll}
\multicolumn{1}{c}{} & \multicolumn{1}{c}{\rotatebox{45}{Expedia}} & \multicolumn{1}{c}{\rotatebox{45}{FiscalNote}} & \multicolumn{1}{c}{\rotatebox{45}{Instacart}} & \multicolumn{1}{c}{\rotatebox{45}{KAYAK}} & \multicolumn{1}{c}{\rotatebox{45}{Klarna}} & \multicolumn{1}{c}{\rotatebox{45}{Milo}} & \multicolumn{1}{c}{\rotatebox{45}{OpenTable}} & \multicolumn{1}{c}{\rotatebox{45}{Shopify}} & \multicolumn{1}{c}{\rotatebox{45}{Slack}} & \multicolumn{1}{c}{\rotatebox{45}{Wolfram}} & \multicolumn{1}{c}{\rotatebox{45}{Zapier}} \\ \hline
Has privacy policy & \chkmark~\cite{expedia_privacy} & \chkmark~\cite{FiscalNote_privacy_policy} & \chkmark~\cite{instacart_privacy_policy} & \chkmark~\cite{kayak_privacy_policy} & \chkmark~\cite{klarna_privacy_policy} & \chkmark~\cite{milo_privacy_policy} & \chkmark~\cite{opentable_privacy_policy} & \chkmark~\cite{shopify_privacy_policy} & \chkmark~\cite{slack_privacy_policy} & \chkmark~\cite{wolfram_privacy_policy} & \chkmark~\cite{zapier_privacy_policy} \\
Mentions LLM$\ast$ & \crossmark & \crossmark & \crossmark & \crossmark & \crossmark & \chkmark & \chkmark & \crossmark & \chkmark~\cite{slack_privacy_ai} & \crossmark & \crossmark\\
Shares with partners$\ast\ast$ & \chkmark & \chkmark & \chkmark & \chkmark & \chkmark & \chkmark & \chkmark & \chkmark & \chkmark & \chkmark & \chkmark
\end{tabular}
\caption{Privacy policies of third-party plugins for OpenAI's ChatGPT. $\ast$ Including generative AI, chatbot, and other names for LLM. $\ast\ast$ Including affiliates, business partners, and service providers.}
\label{tab:plugin_privacy}
\end{table*}

\subsection{Online LLM Services}
\label{sec:bg_llm}

Cloud-based LLM services, such as OpenAI's ChatGPT\cite{chatgpt_web} and Microsoft's Copilot~\cite{microsoft_copilot_web},
have dramatically improved the accessibility of LLM to regular users
through web interfaces.
These online services enable a personal device to access powerful LLM models hosted on the cloud, which are trained with the entire human knowledge available on the internet~\cite{touvron2023llama}.
Public health~\cite{biswas2023role} and education~\cite{adeshola2023opportunities} are examples of many areas that have benefited from the use of online LLM services.

The LLM service providers specify their data collection and usage practices in their privacy policies.
For example, Microsoft's privacy statements indicate that Bing's generative AI feature, namely Copilot, follows the same data collection policy as Bing web search, which may include user's ``searches or commands'' ``(in the form of text, voice data, or an image),'' and the de-identified data are shared with ``selected third parties''~\cite{microsoft_privacy_statement}.

OpenAI's privacy policy states that usage data may be collected for improvement, research, development, misuse prevention, and other business purposes, and the data may be provided to third parties, such as ``providers of hosting services, customer service vendors, cloud services, email communication software, web analytics services, and other information technology providers,'' and business account administrators~\cite{openai_privacy_policy}.

Likewise, Anthropic (which runs Claude.ai) states in its privacy policy that it collects ``Inputs and Outputs'' of its LLM models for service improvement, research, and marketing purposes, and the data may be retained ``for as long as reasonably necessary''~\cite{claude_ai_privacy}.

\subsection{Benefits and Risk of Prompt Collection}
\label{sec:bg_risk}

Users are increasingly concerned about their privacy in using online services.
Microsoft privacy report indicates that 14 million users have visited Microsoft privacy dashboard in a six-month period in 2023, representing a 21\% increase from the previous reporting period~\cite{microsoft_privacy_report}.
The industry and academia have proposed various solutions to mitigate privacy issues in cloud-based Artificial Intelligence (AI), such as differential privacy~\cite{abadi2016deep} and federated learning~\cite{mcmahan2017communication}.
However, online LLM services are not able to fully leverage these solutions due to the fact that they require user prompts in plaintext for training and providing third-party services.

Well-structured prompts
provided by users
are valuable resources.
~\cite{zamfirescu2023johnny, lin2024write}.
As stated in their privacy policies, LLM service providers collect prompts for service improvement and research purposes~\cite{microsoft_privacy_statement, openai_privacy_policy}.
Because of the black-box nature of LLM service providers, the usage of prompts is largely unknown to users.
Past incidents have highlighted that
user queries made available to research may be leaked inadvertently.
In 2006, 20 million
AOL web search queries, intended for research purposes,
were accidentally released to the public, resulting in the identification of individuals in real life~\cite{nyt_aol_leak}.

One of the widely-adopted
usage of prompts is to improve the quality of subsequent LLM responses for the same user.
Fine-tuning and in-context learning are commonly used techniques to improve the relevance of LLM responses to user's previous inputs~\cite{touvron2023llama}.
The prompts used to fine-tune models are kept and processed by LLM service providers' infrastructure.
While these models are not directly exposed to other users, there is a risk of sensitive information being leaked through data breaches or novel attacks that exploit shared infrastructure.

There is a
increased risk of personal information leakage through
adversarial ML attacks if
other users can interact with the fine-tuned model.
Model inversion attacks are a type of privacy attack that aims to recover the training data of a machine learning model by querying the model with maliciously crafted inputs~\cite{fredrikson2015model}.
Prompts provided for training may contain PIIs, medical records, legal documents, and other sensitive information, which can be recovered by adversaries.
We discuss our threat model in Section \ref{sec:model}.

Besides advanced ML model inversion attacks, data breaches are another major concern of prompt collection.
Recent server-side data breaches~\cite{nyt_yahoo_breach, fortune_equifax, nyt_ms_breach} have exposed a massive amount of sensitive personal information, including social security numbers, credit card numbers, and personal history.
A security breach in 2023 at OpenAI exposes internal data~\cite{openai_data_breach}.
These incidents highlight the risk of retaining PIIs and other sensitive information
on servers.

Even legitimate uses of sensitive personal information can be detrimental to user's interest.
Personal web browsing and searching history are commonly used to provide personalized advertisements, and they may be used for price discrimination~\cite{esteves2016competitive,mauring2021search} and even insurance decisions~\cite{online_privacy_insurance_company}.
As a result, users are increasingly concerned about sharing personal information with third parties~\cite {naeini2017privacy}.

Fortunately, it is not necessary for users to provide PIIs to use online LLM services, and users can avoid sharing sensitive information by crafting prompts carefully.
However, this requires users to carefully examine every prompt that they provide to LLM service providers, which is not a scalable solution for users to protect their privacy.

\subsection{Third-Party Extensions to AI Services}
\label{sec:bg_extension}

\subsubsection{Voice Assistant Extensions}

AI-based online services, such as voice assistant and LLM, on their own are limited in functionalities.
Third-party extensions are needed to enhance the capabilities of these services.
Third-party extensions running on top of Amazon's voice assistant, Alexa, are called \textit{skills}~\cite{alexa_skill}.
Skills have been widely adopted for Internet of Things (IoT) control, entertainment, and productivity~\cite{alexa_skill}, but they have also raised privacy concerns.

The poor privacy practice of voice assistant extensions has been widely reported: according to a study in 2021 by Lentzsch et al., only 24.2\% of all skills provide a privacy policy link, and approximately ``23.3\% of privacy policies are not fully disclosing the data types associated with permissions requested by the skill''~\cite{lentzsch2021hey}.
This observation
corroborates the findings of a 2017 study by Alhadlaq et al., that 25\% of around 10,000 skills have no privacy policy~\cite{alhadlaq1902privacy}.
Although the number of skills has rapidly increased from 135 in 2016 to about 125,000 in 2021~\cite{edu2022measuring}, as the statistics from \cite{lentzsch2021hey} and \cite{alhadlaq1902privacy} show, the availability of privacy policies of Alexa skills has not improved from 2017 to 2021.

To address the privacy concerns in third-party skills, since 2023, Amazon has required all skill developers to provide a privacy policy link before their skills can be published at the Alexa Skill Store~\cite{alexa_privacy_url}.
However, privacy policies may not be sufficient to protect users' privacy, as many of them are not accurate~\cite{lentzsch2021hey}, and the actual enforcement of privacy policies is unclear.
Megamind~\cite{seyedtalebi2021_2} is a system solution deployed at the voice assistant devices to enforce privacy control before the voice data is sent to the cloud.
It provides strong privacy guarantees by ensuring that the voice data is processed locally and only necessary information is sent to the cloud.

\subsubsection{LLM Plugins}
\label{sec:bg_llm_plugin}

Parallel to Alexa skills,
third-party plugins are increasingly popular in online LLM services to enable LLM to interact with the real world.

As shown in Figure \ref{fig:llm_services}, online LLM service providers send live or stored LLM interactions to third-party providers, so that they can use the prompts and responses to provide services
~\cite{gpt_plugins, copilot_plugins}.
LLM plugins share some similarities with voice assistant skills, but their functionalities are more diverse, exposing users to a wider range of privacy risks.
Although both Microsoft Copilot and OpenAI ChatGPT's plugin features are currently in their early stages,
they are expected to gain popularity in the near future.

One special plugin is the ChatGPT retrieval plugin.
As mentioned in the plugin's security statements,
instead of retrieving information from the Internet, the plugin searches ``a vector database of content'' to ensure that ``it doesn’t have any external effects''~\cite{gpt_plugins}.
This security consideration acknowledges and mitigates the risk of exposing user's prompts to an Internet search engine, but for third-party plugins, the risk is unchartered territory.

Microsoft Copilot has required third-party providers to provide privacy policies for their plugins~\cite{copilot_plugins}.
However, as we have seen in the case of voice assistant skills, mandating third-party providers to provide
privacy policies may not be sufficient to protect sensitive information in user's prompts.
Indeed, according to our study, although all plugins showcased on ChatGPT Plugins page~\cite{gpt_plugins} have privacy policies, only 27.3\% of plugins' privacy policies mention LLM.
All the surveyed privacy policies further mention that they share user data with their business affiliates, partners, or service providers (which are fourth parties from the perspective of LLM users).
We show the statistics in Table~\ref{tab:plugin_privacy}.

\section{Threat Model}
\label{sec:model}

We consider users to be honest, are using a not compromised system and browser to access online LLM services,
and are willing to share their prompts with LLM service providers and certain third-party providers (through LLM extensions).
However, they are concerned about personal privacy and do not want to burden the providers with the responsibility of protecting their sensitive information in the prompts.

We consider LLM service providers to be only trusted for correctly fulfilling user's requests, but not trusted for withstanding malicious attacks.
Under such assumption, LLM service providers communicate with users and third-party providers through encrypted channels, and use industrial standard technologies to protect user's data from unauthorized access.
However, LLM providers may collect and use user's prompts for business-related purposes, such as improvement and research, which may unintentionally disclose sensitive information through data breaches or ML model inversion attacks~\cite{fredrikson2015model}.

We assume the same threat model for third-party providers as LLM service providers,
but they are at a higher risk due to the fact that
they may have various
privacy and security policies and practices.

\section{Motivation and Design Goals}
\label{sec:motivation}

\subsection{Straw-Man Solution 1: Local LLM Inference}
\label{sec:motivation_local_llm}

One obvious solution to the privacy concerns of online LLM services is to perform all LLM inference locally on user's device.
Indeed, local LLM inference has been made possible by open-source LLM models~\cite{touvron2023llama}.
WebLLM leverages WebGPU to accelerate LLM inference using local CPU or GPU through a web browser~\cite{webllm}.
Likewise, TinyChat is a framework for local LLM inference on edge devices, such as robots and in-car systems~\cite{tinychat}.

However, local LLM inference is not comparable to cloud-based LLM services in terms of performance and scalability.
The training cost of OpenAI's GPT-4 model costs over 100 million dollars~\cite{wired_gpt4}, and so does the training cost of Google's Gemini Ultra model~\cite{stanford_ai_report}.
Inference of these models requires an (undisclosed) large amount of computational resources, which are not possible to be run on user's devices.

\subsection{Straw-Man Solution 2: Cloud-based PII Redaction}

Another solution to address the privacy concern is to perform PII redaction and other sensitive information sanitization on the LLM service providers' side.
The LLM service providers can deploy black-box algorithms to remove sensitive information from users' prompts before they are processed by LLM models.
The black-box algorithm may even utilize an LLM model to identify sensitive topics in the prompts.

However, this approach creates a trust issue between users and LLM service providers.
As mentioned in our threat model in Section~\ref{sec:model}, users may not trust LLM service providers to protect their privacy, as the providers have different privacy policies, and may even have incentives to retain sensitive information.

\subsection{Research Challenges and Design Goals}

\Sname's goal is to
address the privacy concerns in submitting sensitive information to online LLM services.
Compared with running LLM inference locally, cloud-based LLM services are necessary for users to access prohibitively large models and to leverage the latest model updates.

It might be tempting to perform PII redaction on the LLM service providers' side, but this approach is not ideal.
The LLM service provider may not have the incentive to
protect users' privacy as much as the users themselves.
Also, users may not trust LLM service providers to use black-box algorithms to
protect their privacy.
We identify the following research problems, which are also design goals, for \Sname to address:

\paragraph{\textbf{Goal 1: Comprehensive Privacy Protection}} The primary goal of \Sname is to provide comprehensive privacy protection for cloud-based LLM services, by filtering out sensitive personal information with or without a fixed pattern.

\paragraph{\textbf{Goal 2: Client-Side Filtering}} \Sname should run entirely on user's device without using any cloud-based services, so that user's sensitive information never leaves their devices. This includes the LLM inference for topic identification and any other filtering processes.

\paragraph{\textbf{Goal 3: Lightweight and Efficient}} The system design and implementation should not introduce significant performance overhead to the user's device, and should not burden users with complex configurations or software installations.
The best user experience is achieved when \Sname is readily deployable on a web browser as a browser extension.

\paragraph{\textbf{Goal 4: Compatibility}} The system architecture should be compatible with existing online LLM services.
\Sname should not require any modification to the existing web interfaces of online LLM services.
Other than privacy protection, the user interaction with the LLM services should remain the same.

\paragraph{\textbf{Goal 5: User Awareness and Control}}
If a prompt contains sensitive topics but not specific keywords or patterns, the user should be informed of the potential privacy risks.
The warning should be clear and informative to raise user awareness of the privacy risks.
Users should have the ability to customize the filtering rules to include user-specific patterns and topics that they do not want to share with LLM services.

\section{Design}
\label{sec:design}

In this section, we present the design of \Sname.
We present a system overview, followed by an overview of our three-layered prompt sanitization mechanism.
We then discuss the design decisions and trade-offs in each of the three layers, i.e., rule-based filtering, ML-based named-entity detection, and local LLM-based topic identification.
We also discuss the research challenges in the data preparation for local LLM topic identification, and the user interface design for awareness-raising.
Finally, we present the feature-preserving redaction mechanism in \Sname.

\begin{figure*}
\centering
\includegraphics[width=0.95\textwidth]{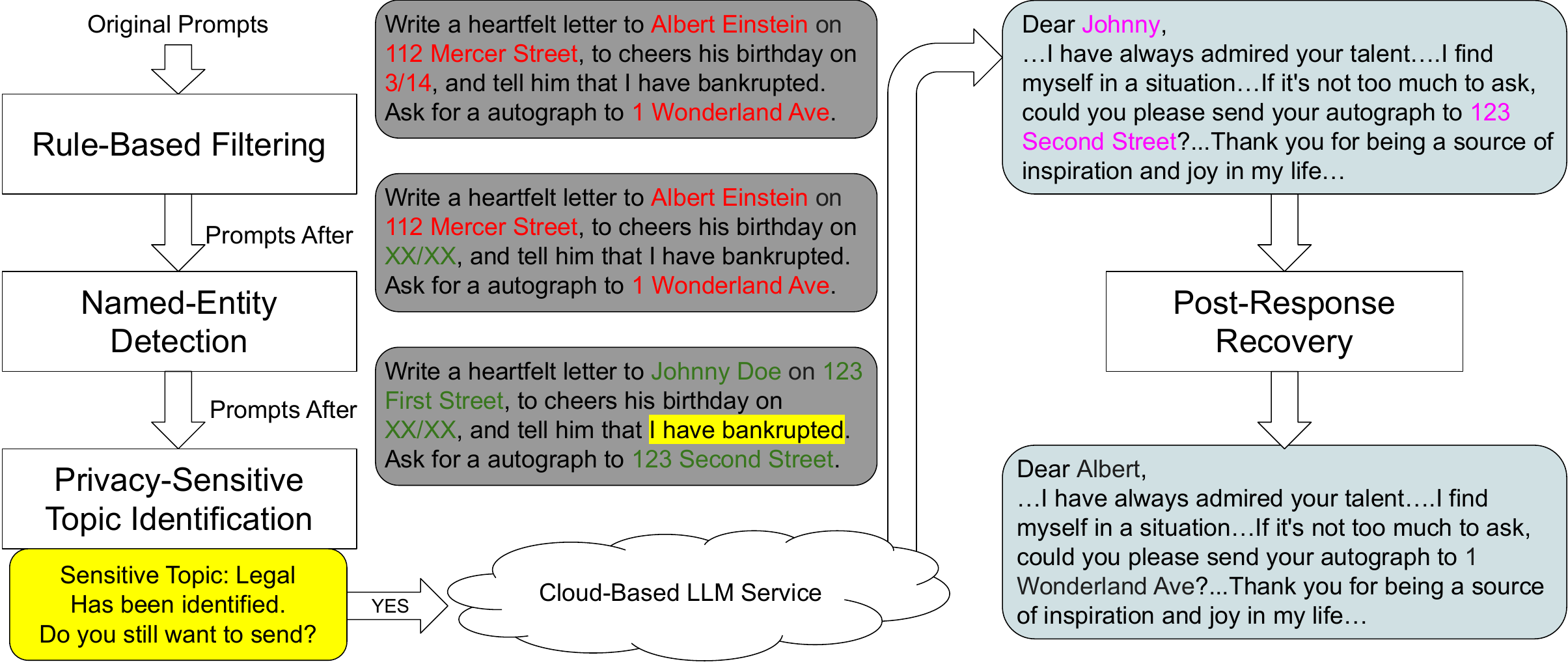}
\caption{Three-layered prompt sanitization in \Sname.}
\label{fig:three_stage}
\end{figure*}

\subsection{System Overview}
\label{sec:overview}

As shown in Figure~\ref{fig:llm_ours}, \Sname examines the prompts provided by users when they interact with online LLM services.
\Sname identifies two types of private information in the prompts: PIIs and sensitive topics.

\Sname redacts PIIs from the prompts by replacing them with unique placeholders
before sending them to LLM services.
By sharing only the placeholders with LLM services, \Sname
ensures that they can still generate relevant responses without knowing the original PIIs.
And when the LLM service returns the responses, \Sname
reverts any placeholders in the response back to their original forms before displaying the responses to the users.

For sensitive topics, such as personal medical and legal information in user's prompts, \Sname identifies them and alerts users to review and confirm before sending them to the LLM service.
The user interface of \Sname clearly indicates the sensitive topics involved in the prompts, and helps users to make informed decisions about sharing them with LLM services (fulfilling our goal of user awareness and control, Goal 5).

As shown in the example in Figure \ref{fig:three_stage}, \Sname
filters out sensitive identifiers, alerts users of sensitive topics, and then forwards the sanitized prompts to an unmodified LLM service pipeline (fulfilling our goal of compatibility with existing LLM services, Goal 4).
As the prompts are sanitized before they leave the user's device, the LLM service provider, plugins, and other third-party providers only receive sanitized prompts and do not have access to the original prompts.

\begin{figure}
\centering
\includegraphics[height=1.5in]{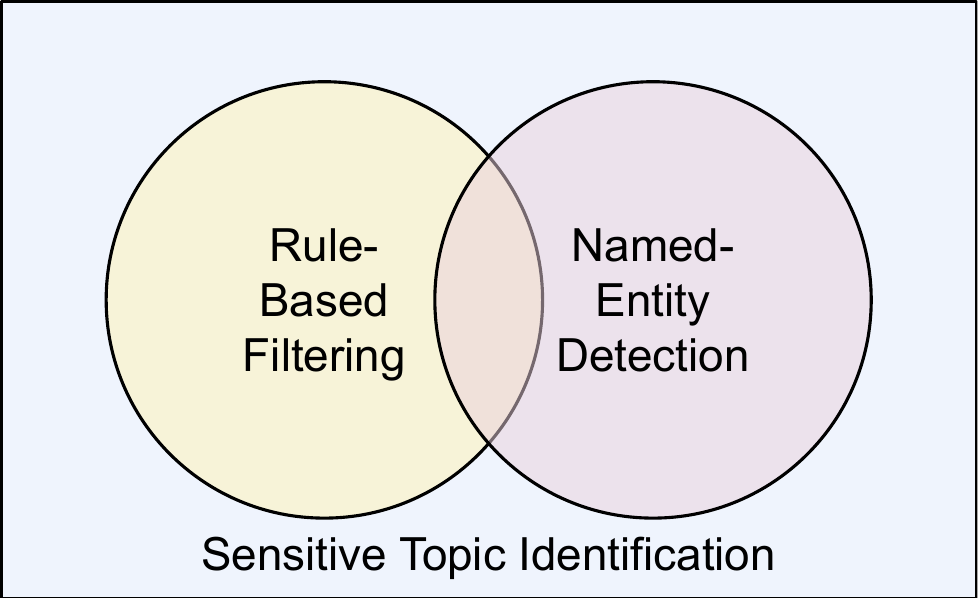}
\caption{Venn diagram of the coverage of three stages of prompt sanitization in \Sname.}
\label{fig:three_stage_venn}
\end{figure}

\subsection{Three-Layered Prompt Sanitization}
\label{sec:three_stage}

\Sname sanitizes the prompts in three separate stages: rule-based filtering, ML-based named-entity detection, and local LLM-based topic identification.
The rule-based filtering identifies fixed patterns, as well as user-defined patterns and keywords.
The ML-based named entity detection identifies names and other identifiers that do not follow fixed patterns.

The combination of these three stages is based on a sound rationale:
As shown in Figure \ref{fig:three_stage_venn}, rule-based filtering and named-entity detection cover a wide range of sensitive PIIs and may overlap in identifying known patterns and
named entities of unknown patterns
in the prompts.
However,
a prompt may contain information about certain sensitive topics that are not readily identifiable with fixed patterns, and may not contain any named entities.
Local LLM-based topic identification is designed to catch the information that may be missed by the first two stages.
These three stages are designed to complement each other, fulfilling our design goal of comprehensive privacy protection (Goal 1).

\Sname runs entirely on user's device, and does not rely on any cloud-based services.
All these three stages
are self-contained within our browser extension, fulfilling our design goal of client-side filtering
and lightweight and efficient design (Goal 2 and Goal 3).
Although prompts go through three stages of filtering,
we demonstrate that the performance overhead of \Sname is minimal (14.7\%).
We present a comprehensive evaluation of the effectiveness and performance overhead of \Sname in Section~\ref{sec:eval}.

\subsection{Rule-Based Filtering}

Rule-based filtering is the first stage of prompt sanitization.
It identifies generic patterns that are commonly associated with a person's identity, such as alphanumeric patterns that resemble a phone number, social security number, credit card number, and email address.
Users can further customize the rules in the form of regular expressions to include or exclude specific patterns.
The rule-based filtering also identifies keywords that users provide, which they may consider sensitive, such as their names, cellphone numbers, and other keywords that they do not intend to share with LLM services.

\subsection{Named-Entity Detection}
\label{sec:design_named_entity}

Named entities are words that refer to a certain entity, such as a person, a location, an organization, or a miscellaneous entity.
Miscellaneous entities can be any entities that do not fall into a specific category.
For example, a product name or an event name is a miscellaneous entity.
Named-entity cannot be readily identified by rule-based filtering, as they can be any word, and may not follow a specific pattern.
Therefore, to complement the rule-based filtering, we adopt an ML-based named-entity detection model to identify them in the prompts.
A named entity detection model is trained on top of a large dataset of named entities, and the accuracy of the model depends on the quality and quantity of training data.

Named entity detection is an evolving field in natural language processing, and there are many pre-trained models available~\cite{microsoft_presidio, souza2019portuguese, chang2021chinese, bert}.
One research challenge is to create an architecture that can plug and play different named entity detection models.
To address this challenge, we design a modular architecture using Transformers.js \cite{transformers.js} to load a pre-trained model in our browser extension.
We leverage the common design of named entity detection models, where the model takes a string as input and returns a list of identified entities in the string.
We tailor \Sname's programming interface to be compatible with the common design, so that \Sname can easily adopt different named entity detection models that fit user's need and language preference.

\subsection{Sensitive Topic Identification}
\label{sec:design_llm_topic}

The topic identification is the final stage of prompt sanitization.
The reason behind this stage is that prompts may reveal sensitive personal situations without containing any keywords, patterns, or named entities.
One such example we raised in Section~\ref{sec:intro} is that a user may ask questions about their medical symptoms without mentioning any specific medical jargon.
As mentioned in Section~\ref{sec:bg_risk}, this information will be shared with third-party providers and may be used against the user's interest, such as price discrimination
~\cite{esteves2016competitive,mauring2021search} and insurance decisions~\cite{online_privacy_insurance_company}.

Our design addresses this issue by deploying a local LLM model running entirely in a browser extension to identify sensitive topics in the prompts.
\Sname adopts WebGPU-based LLM inference architecture~\cite{webllm} to accelerate the LLM inference.
Users have full control over the LLM model and can choose to use various pre-trained models of different model sizes and context lengths
available online that fit their computers' configurations.

We show the template of the local LLM query for topic identification below,

\begin{tcolorbox}
\textbf{Local LLM Query for Topic Identification:} Below is a prompt. Answer only yes or no based on if the prompt contains \textit{TOPIC} information. "\textit{DATA}"
\end{tcolorbox}

An actual query replaces the \textit{DATA} field with the user's prompts, and the \textit{TOPIC} field with the enumeration of sensitive topics.
\Sname
has a built-in list of sensitive topics,
and it allows users to add any topics they consider sensitive by simply adding the names of topics to an input field in our user interface.
These topics will be automatically converted to additional
LLM queries to identify the requested topics in the prompts.

We understand that a LLM model may not be perfect in identifying sensitive topics, and may generate false positive and false negative results.
However, as we observed in our evaluation (Section~\ref{sec:eval}), larger LLM models are effective in identifying sensitive topics.
For example, the accuracy is relatively high (66.8\%) using the Llama 2 model, and even more accurate (96.6\%) using the Llama 3 model.
Indeed,
according to our experiments,
recent laptops are capable of running the Llama 2 7B model at 12.88 tokens per second, the Llama 3 7B model at 10.52 tokens per second, and the Tiny Llama 1.1B model at 50.73 tokens per second.
Users can choose to use a smaller pre-trained LLM model if hardware resources are limited, but the trade-off is that a bigger model is more effective in identifying sensitive topics.

\Sname leverages Web LLM~\cite{webllm}, which uses browser support for WebGPU~\cite{webgpu} to load pre-trained LLM models in our browser extension.
Our programming interface sends queries to a local LLM model, and collects the responses to see if the prompts contain sensitive topics.
Similar to the design philosophy of our plug-and-play named entity detection models, \Sname is designed to be
generic and compatible with any model on the Web LLM platform~\cite{webllm}, including the Llama 2, Llama 3, TinyLlama, RedPajama, and other models.

\subsection{Data Preparation for Local LLM Topic Identification}

We face a research challenge as we feed our query to the local LLM model.
Our query template (as shown in Section \ref{sec:design_llm_topic}) is a combination of our request to identify a certain topic in the prompt, and the prompt itself.
The prompt may contain questions, statements, data, and other information, that confuses the LLM model.
This challenge has been documented by ~\cite{zverev2024can} where the authors show that LLMs cannot distinguish between user's questions and attached data, and therefore often generate irrelevant responses.
The suggested solution in ~\cite{zverev2024can} is to improve the LLM model in explainability, but this is a long-term research direction, and not applicable to our current work as we are using a pre-trained model.

However, we discover a lightweight solution to our unique use case of topic identification.
Psycholinguistic research has shown that the informativeness of a sentence is mainly carried by the nouns (including Direct Object (DO) nouns)~\cite{lyu2019neural}.
Inspired by this observation, we extract the nouns from the user's prompt, and use them as the data field of our local LLM query.
This design decision not only resolved the model confusion problem mentioned earlier, but also reduced the token count of the prompt and improved the processing time of the local LLM model.

In our design, we also address the potential issue of long prompts.
As LLM models have token limits, and online LLM services may have a bigger limit than local LLM models, we truncate the prompt to the maximum token limit of the local LLM model, and query the local LLM model for one truncated section of the prompt at a time.

\begin{algorithm}[h]
\caption{Local LLM Topic Identification}
\label{alg:llm_topic}
\SetKwFunction{checkTopic}{checkTopic}
\SetKwProg{Fn}{Function}{:}{}
extract nouns\;
count currTokens\;
update totalTokens\;
\If{totalTokens > limit}{
divide nouns into chunks of maximum token limit\;
\For{each chunk}{
reload model\;
reset totalToken\;
\checkTopic{$chunk$}\;
}
}
\Else{
\checkTopic{$nouns$}\;
}
return response\;
\Fn{\checkTopic{$nouns$}}{
\For{each topic}{
\If{totalTokens > limit}{
reload model\;
}
send $nouns$ to local LLM\;
wait for response\;
update totalToken\;
clear chat history\;
\If{sensitive topic detected}{
return response\;
}
}
}
\end{algorithm}

\subsection{Awareness-Raising User Interface}

\Sname warns users about privacy risks when their prompts contain PIIs or sensitive topics.
The warning message
highlights the private and sensitive information detected in the prompt, and provides a reference to the online LLM service's privacy policy.
The warning message is designed to be clear and informative,
and its user interface ensures that users can
quickly understand the privacy risks if they decide to share the prompts with LLM services.
If a user decides to proceed, they can click a button to acknowledge the warning and send the prompt to the LLM service.

\subsection{Feature-Preserving Redaction}
\label{sec:design_redaction}

\Sname replaces PIIs detected by pattern-based filtering and named-entity detection with unique placeholders.
For person names and other named entities, pseudo names are generated using Faker.js \cite{faker.js}.
\Sname maintains a mapping of the original PIIs and the placeholders, allowing it to revert the placeholders to the original PIIs when a response is received from the LLM services.

This design has several benefits:
First, it ensures that LLM services can still generate relevant responses without knowing the actual PIIs.
Second, the redaction process is behind the scenes and does not affect the user experience.
Because of the user-transparent design of the PII redaction,
(although the false positive rate of named-entity detection is relatively low (13.3\%)),
\Sname tolerates false positives in the redaction process as all placeholders are reverted to their original forms in the responses displayed to users.
Users can still interact with the LLM services as they normally would, and the responses are displayed with their original PIIs as if the redaction process never happened.
\section{Implementation and Prototype}
\label{sec:impl}

\subsection{\Sname Browser Extension}

\Sname is implemented as a browser extension for Google Chrome version 124.
The extension is written in TypeScript, HTML, and CSS in approximately
one thousand
lines of code.
The browser extension features an icon in the browser toolbar, which users can click to open \Sname's configuration panel.
The configuration panel features personal information setup, regular expression customization, sensitive topic configurations, model selections for local LLM,
and local LLM and named entity detection settings.

Although \Sname's design applies to other browsers and online LLM services, we focus on Google Chrome and OpenAI's ChatGPT in our prototype implementation.
The extension activates itself when user visits the ChatGPT website, and it analyzes user's prompts using the three-layered filtering mechanism (Section~\ref{sec:design}).
We implement all three layers of prompt sanitization in TypeScript in the browser extension, and therefore, there is no need for any software installation or cloud-based services.

\subsection{Rule-Based Filtering}

The filtering rules are implemented as keyword search and regular expression match in TypeScript.
Starting with common PII patterns, such as email addresses, phone numbers, and passwords,
we iteratively improve the regular expression patterns to include more PII patterns with the help of LLM.
Note that we use LLM merely as a development tool
and do not introduce LLM in the rule-based filtering process.

We ask LLM to generate worldwide PII examples,
evaluate our regular expression patterns against the generated strings, and add the missing patterns to our regular expression pool.
We repeat this process until we are confident that our regular expression patterns cover most PII patterns.
Our regular expression pool currently consists of 10 patterns.
Each pattern covers several formats of a specific PII such as various formats of phone numbers.
Users may include their own keywords or patterns in the filtering rules through the configuration panel.

\subsection{Named-Entity Detection}

We use bert-base NER \cite{bert_base_NER} which is a fine-tuned BERT model \cite{bert} for named-entity detection.
The model has 110 million parameters and is trained on CoNLL-2003 Named Entity Recognition \cite{tjongkimsang2003} dataset,
which includes 6600 person names, 7140 locations, 6321 organizations, and 3438 miscellaneous examples.
Although the model is implemented for TensorFlow \cite{tensorflow} toolchain, we utilize Trasnformers.js \cite{transformers.js} to load the model from within our browser extension.
\Sname sends user's prompts to the local Transformers.js interface, and synchronously receives results from the model in the form of a list of recognized entities in the string.
Each of the recognized entities is tagged with its type and a score indicating the model's confidence in the recognition.
This score provides an opportunity for future enhancement, such as only reporting recognized entities above a certain threshold.
\Sname provides a default score threshold of 90 out of 100, which can be adjusted by users.

Through \Sname's configuration panel, users can turn on or off the named-entity detection feature.
As discussed in Section~\ref{sec:design_named_entity}, the named-entity detection model can be easily replaced with other models, as long as the model is compatible with the Transformers.js library.

\subsection{Local LLM-based Topic Identification}

\Sname uses a local LLM model for topic identification.
We implement local LLM inference by leveraging the WebLLM framework~\cite{webllm}.
We port the WebLLM framework to our browser extension.
The WebLLM framework provides high flexibility in changing the LLM model, therefore, it makes switching between different LLM models easy in \Sname.
Users can select from a few LLM models on the configuration page of \Sname based on their needs.
The local LLM model \Sname uses is the Llama 3 model with 8 billion parameters.
As shown in Section \ref{sec:eval_topic}, we choose the Llama 3 model for its balance between performance, accuracy, and resource consumption.

\Sname sends user's prompts to the local LLM model through Web LLM API and synchronously reads a binary result from the model which indicates whether the prompt contains privacy-sensitive topics.
Users can optionally turn on or off the topic identification feature and the default Llama 3 model can be replaced with other models through our configuration panel.
Indeed, before the Llama 3 release on April 18, 2024~\cite{llama3}, the best trade-off among our criteria was the Llama 2 model, which we used in our early prototype development.
With less than 10 lines of code modification and approximately 15 minutes of human effort, we successfully switched to the Llama 3 model.

During our development, we notice that the detection accuracy degrades significantly when we ask the LLM model to identify more than one topic, such as both medical and legal topics.
We show the results of this experiment in Table~\ref{tab:multiple_vs_single}.
To address this issue, we perform topic identification for each topic separately.
For example, if a user asks if a prompt contains both medical and legal topics, we first ask the local LLM model to identify medical topics, and then ask it to identify legal topics.
To avoid previous inputs from influencing the detection in the subsequent prompt, \Sname clears the chat history after each run of topic identification.

\begin{table}[h]
\centering
\begin{tabular}{|c|c|}
\hline
Single Topic & Multiple Topics \\ \hline
96.6\% & 32.4\% \\ \hline
\end{tabular}
\caption{Detection accuracy of single topic and multiple topics.}
\label{tab:multiple_vs_single}
\end{table}

\subsection{User Interface Integration}

We implement \Sname to tailor the user interface of OpenAI's ChatGPT.
Our browser extension locates the prompt input box in ChatGPT web interface and analyzes user's prompts before they are submitted.
Due to the limitations of Chrome extension API, we cannot intercept the input prompt right before it is sent to the LLM service provider through Enter keystroke or submission button click.
We have two options to address this Chrome extension limitation:
1) we can analyze the prompt on every keystroke, or 2) we can introduce a new keystroke combination to analyze the prompt, for example, Control-Enter.
In our prototype implementation, we opt for the second option, as it is more user-friendly and
avoids the performance overhead of analyzing the prompt on every keystroke.
One may argue that users may forget to press Control-Enter before submitting prompts and we acknowledge this concern.
However, we can introduce enforcement from \Sname by blocking the prompt submission until user presses Control-Enter.
Chrome extension API allows us to intercept the submission event and prevent the prompt from being sent to the LLM service provider by using our own rulesets.
However, we can only block the submission event. We cannot modify the content of the prompt before it is sent to the LLM service provider.

\begin{figure}
\centering
\includegraphics[width=0.4\textwidth]{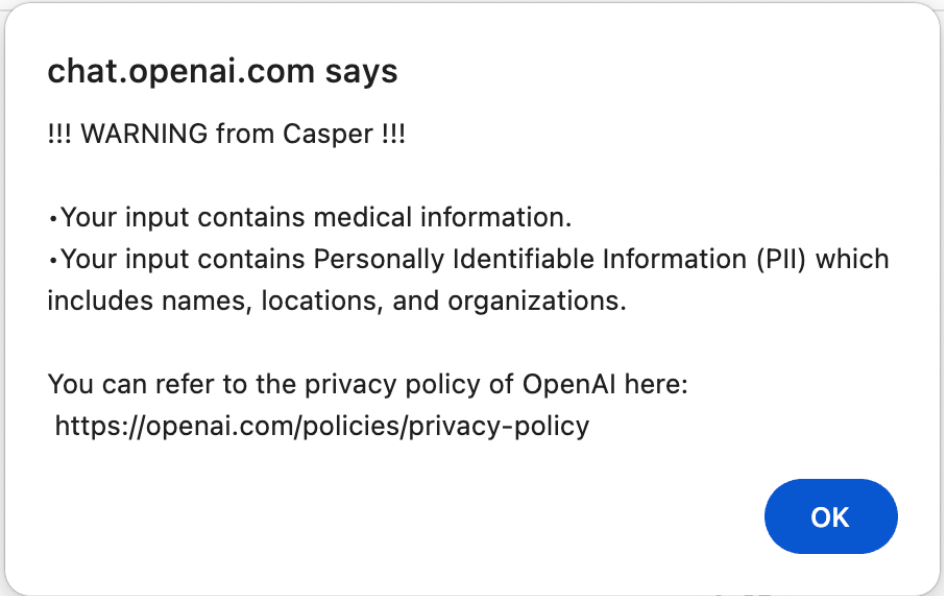}
\caption{Popup message box that indicates private and sensitive information found by \Sname.}
\label{fig:casper_alert}
\end{figure}

When the user presses Control-Enter, \Sname analyzes the prompt using the three-layered filtering mechanism.
The PIIs detected in the first two layers are automatically redacted using the methodology described in Section~\ref{sec:design_redaction}.
Specifically, we use Faker.js library to generate unique dummy placeholders for each PII detected, and store the mapping between them in the browser's local storage via Chrome extension storage API.
For the third layer, we query the local LLM model to identify privacy-sensitive topics.
Finally, if \Sname detects any privacy or sensitive information in any of the three stages, \Sname will use \textit{window.alert()} method to display a warning window notifying the user of the privacy risk.
As shown in Figure~\ref{fig:casper_alert},
the warning window is a popup message box that appears on top of the ChatGPT web interface.
It contains a warning message indicating the topic and privacy risk and a button for the user to acknowledge the potential information leakage.
Users will not be able to proceed with the prompt submission until they acknowledge the warning message.
For a more streamlined user experience, users can disable alerts for PII detection and let them be redacted automatically.
However, privacy-sensitive topic alerts are always enabled.

Once the response from the online LLM service is received, we will be able to replace the placeholders with original PIIs using the mappings stored in the local storage.
Due to the long latency and low throughput nature of the LLM responses, we cannot automatically revert the placeholders back to the original PIIs upon receiving the responses.
Furthermore, if such automatic conversion is implemented, it will introduce a large overhead since we have to perform searches as soon as partial responses are received.
\Sname by default displays the original response from the LLM service provider, and users can press Control-Shift to reveal the reverted response.
Similar to the keystroke needed for triggering prompt analysis, the keystroke is a convenient way to interact with \Sname without introducing performance overhead.

\section{Evaluation}
\label{sec:eval}

We evaluate the effectiveness of \Sname in protecting user privacy and its performance overhead.

\subsection{Evaluation Setup}
\label{sec:eval_setup}

We evaluate \Sname on both Apple MacBook Air and MacBook Pro with an 8-core CPU, 7-core GPU, and 8GB of unified memory and
10-core CPU, 14-core GPU, and 32GB of unified memory, respectively.
Both MacBooks run on macOS Sonoma and Chrome version 124.
Our Internet connection has a download speed of 300mbps and an upload speed of 250mbps.
To demonstrate that even a consumer-grade laptop with 8GB of memory can perform named entity detection,
we utilize MacBook Air for the evaluation of named entity detection, and MacBook Pro for the evaluation of sensitive topic identification.

We automate the testing of \Sname by using TypeScript that automatically processes CSV files that contain synthesized prompts.
We process each row of data as if we receive the prompt directly from the user so there won't be any discrepancy between the testing and real-world scenarios.
As the file is uploaded to our extension, it will trigger our three-staged prompt sanitization and save the results to another CSV file, which
will be automatically downloaded to our local machine for further analysis.

\subsection{Named-Entity Detection}
\label{sec:eval_named_entity}

We evaluate \Sname on 1000 synthesized prompts with named entities and another 1000 prompts without named entities.
We use OpenAI's GPT-4 Turbo model to generate the corpus of prompts for use in our evaluation.
We first compose a list of named entities and then use the following query for GPT-4: \textit{``Generate only one thoughtful question (to ask chatgpt) using one of the personal information (use name or address or company) supplied for each row of the data.
Process all the rows and generate a csv file for the output. There needs to be a variety of questions. No duplications at all''}, with respect to each row of the named entities in the
CSV dataset that we have provided to GPT-4 Turbo model to process.
For the prompts without named entities, we use the following query: \textit{``Generate 200 random questions (does not contain any personally identifiable information) to ask chatgpt''}.
The reason why we query for 200 prompts at a time is due to the limitation of the GPT-4 Turbo model, which might not generate the prompts as requested if the amount of prompts is too large.
We then combine the two sets of prompts to be used in our evaluation,
using our named entity dataset to mark the ground truth.
One example of the synthesized prompts with named entities is \textit{``What are some networking tips for professionals in Wright-Sweeney?''} and without named entities is \textit{``What causes lightning and thunder?''}.

From the evaluation results, it is clear that \Sname can effectively detect named entities in the prompts since the true positive and true negative rates are relatively high as shown in Table~\ref{tab:named_entity_confusion_matrix}.
The false negative in the results mainly comes from the synthesized prompts that are poorly formed using the named entities provided, such as \textit{``Could you list famous landmarks or events associated with the provided address location?''}.
The synthesized prompts use the phrases \textit{``provided information''} instead of the actual information which makes it difficult for \Sname to detect the named entity.
The false positive rate is slightly higher than the false negative rate because some of the synthesized prompts without named entities use names of countries in the prompts, such as ``What is the capital of France?''.
However, this is partly due to the fact that the query we asked GPT-4 Turbo model to generate the prompts without named entities is not accurate enough
and the false positive rate can be reduced further by refining the query such as \textit{``Generate 200 random questions that do not contain any named entities''} instead of \textit{``Generate 200 random questions (does not contain any personally identifiable information) to ask chatgpt''}.
However, we prefer to keep the query as is to demonstrate the robustness of \Sname in a real-world scenario.

\begin{table}[h]
\centering
\begin{tabular}{|c|c|c|}
\hline
\textbf{} & Predicted Positive & Predicted Negative \\
\hline
Actual Positive & 98.5\% & 1.5\% \\
\hline
Actual Negative & 13.3\% & 86.7\% \\
\hline
\end{tabular}
\caption{Evaluation results of named entity detection.}
\label{tab:named_entity_confusion_matrix}
\end{table}

\subsection{Sensitive Topic Identification}
\label{sec:eval_topic}
To evaluate the accuracy of \Sname's LLM-based sensitive topic identification, we use 1000 synthesized prompts (500 with medical topics and 500 with legal topics) and another 1000 ones without any medical and legal topics.
Similar to the named-entity detection evaluation, we generate these prompts with OpenAI's GPT-4 Turbo model.
We generate the synthesized prompts using the following query \textit{``Generate 1 thoughtful question (to ask chatgpt) for each of these 200 medical conditions listed below:''}, and automatically label the prompts as privacy-sensitive or not based on the query.
The same approach is used for legal topics.
For the prompts without sensitive topics, we use the query \textit{``Generate 200 random questions to ask ChatGPT (non-medical and non-legal questions)''}.
One example of the synthesized prompts with medical topics is \textit{``What are the most effective treatments for adult acne compared to teenage acne?''} and without sensitive topics is \textit{``What are the top 5 most beautiful beaches in the world?''}.
All the synthesized prompts are then combined with their relative medical/legal topics to form the evaluation dataset.

We evaluate the detection rates of medical and legal topics separately, and compare the detection rates between Llama 2 and Llama 3.
Llama 3's detection rate in medical topics is 29.8\% higher than Llama 2.
For medical and legal topics, \Sname achieves a detection rate of 96.6\% and 83.2\% respectively with Llama 3 as shown in Table~\ref{tab:medical_confusion_matrix} and Table~\ref{tab:legal_confusion_matrix}.
Some of the terms in the prompts used to evaluate medical topic identification are ambiguous, which results in false negatives.
These ambiguous terms are incorrectly interpreted as non-medical topics, such as autism, down syndrome, and hearing loss.
We have no information on how Llama 3 is trained in identifying medical topics.
However, from our analysis of the results, the terms that are not identified as medical topics are often chronic conditions that do not require immediate medical attention.
For the false positive rate in medical topic identification, some of the prompts stand between medical and non-medical topics, such as ``What are the benefits of aromatherapy?''.
Since the prompt generation and sensitive topic detection are done by two different models, the differences in the models' understanding of a certain sensitive topic can lead to false positives and false negatives.
This is especially true for legal topics, where the false positive and false negative rates are both higher than those in medical topics as shown in Table~\ref{tab:legal_confusion_matrix}.

\begin{table}[h]
\centering
\begin{tabular}{|c|c|c|}
\hline
\textbf{} & Predicted Positive & Predicted Negative \\
\hline
Actual Positive & 96.6\% & 3.4\% \\
\hline
Actual Negative & 4.2\% & 95.8\% \\
\hline
\end{tabular}
\caption{Evaluation results of medical topic identification by Llama 3.}
\label{tab:medical_confusion_matrix}
\end{table}

\begin{table}[h]
\centering
\begin{tabular}{|c|c|c|}
\hline
\textbf{} & Predicted Positive & Predicted Negative \\
\hline
Actual Positive & 83.2\% & 16.8\% \\
\hline
Actual Negative & 5.8\% & 94.2\% \\
\hline
\end{tabular}
\caption{Evaluation results of legal topic identification by Llama 3.}
\label{tab:legal_confusion_matrix}
\end{table}

\subsection{Performance Overhead}

\subsubsection{ChatGPT Response Time}

\begin{table}[]
\begin{tabular}{|ll|l|l|l|}
\hline
&            & Medical & Legal & Random \\ \hline
\multicolumn{1}{|l|}{9-11AM} & Latency    &   0.77  &  1.23 &  1.3   \\ \cline{2-5}
\multicolumn{1}{|l|}{}       & Total Time &    17   & 14.68 & 13.68  \\ \hline
\multicolumn{1}{|l|}{2-4PM}  & Latency    &   0.92  &  0.86 & 0.79   \\ \cline{2-5}
\multicolumn{1}{|l|}{}       & Total Time &  11.77  &  10.2 & 9.88   \\ \hline
\multicolumn{1}{|l|}{7-9PM}  & Latency    &   0.63  &  0.65 & 0.66   \\ \cline{2-5}
\multicolumn{1}{|l|}{}       & Total Time &   7.88  &  7.61 & 7.25   \\ \hline
\end{tabular}
\caption{Average response time of ChatGPT without \Sname. The time is measured in seconds. Latency is the time ChatGPT takes to start to respond to our prompt, and Total Time is the time ChatGPT takes to finish responding to our prompt.}
\label{tab:time_overhead_baseline}
\end{table}

\begin{figure}
\centering
\includegraphics[width=0.4\textwidth]{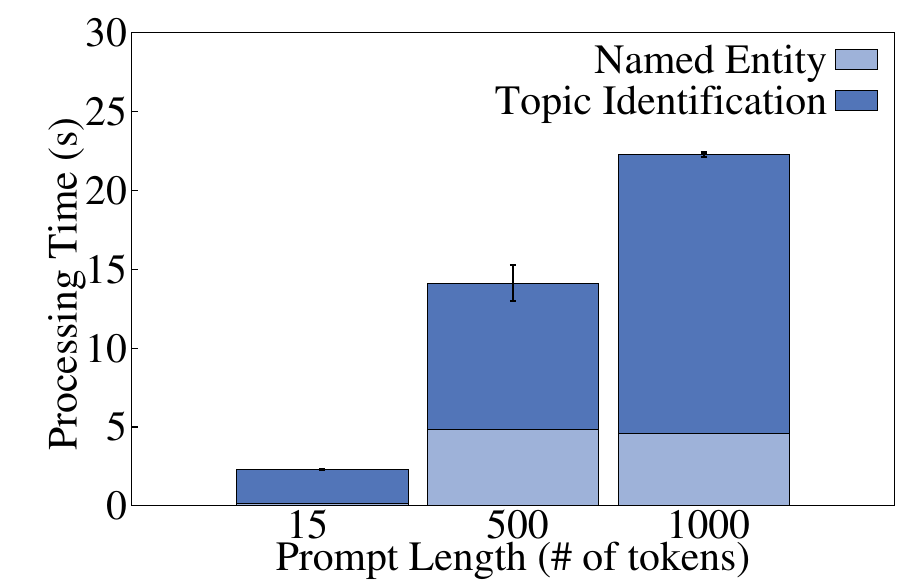}
\caption{Time to process a prompt in each stage of \Sname.}
\label{fig:processing_time}
\end{figure}

We measure the average response time of a normal interaction with OpenAI's ChatGPT web interface.
We
interact with the ChatGPT website at three different times of the day and record the time it takes to respond to our prompts in each topic category.
We repeat each experiment 30 times and calculate the average and standard deviation of (1) the time ChatGPT takes to begin to respond, and (2) the total time ChatGPT takes to complete its response.
We summarize the results in Table~\ref{tab:time_overhead_baseline}.

\subsubsection{\Sname Processing Time}

\Sname processes a prompt in three stages: rule-based filtering, named-entity detection, and sensitive topic identification.
We measure the time it takes to process a prompt in each stage and measure the performance scalability of \Sname by varying the length of the prompt.
For this experiment and the following experiments in this section,
we use the same set of synthesized prompts as in Section \ref{sec:eval_named_entity} and Section \ref{sec:eval_topic} respectively to evaluate the performance overhead of \Sname's named-entity detection and sensitive topic identification.
The processing time of named-entity detection and sensitive topic identification is around 0.2 seconds and 2.15 seconds respectively for prompts that consist of about 15 tokens.
This result is obtained from the experiments that we conducted in Section \ref{sec:eval_named_entity} and Section \ref{sec:eval_topic} which consists of running 4000 synthesized prompts.
To evaluate the performance scalability of \Sname, we have prompts with 3 different lengths: 15 tokens, 500 tokens, and 1000 tokens.
We perform each experiment 3 times and report the average and standard deviation.
Note that we do not separately evaluate the rule-based filtering stage, as the performance overhead in pattern matching is negligible (around 10ms for all prompts).

We report the time breakdown of processing a prompt in each stage in Figure~\ref{fig:processing_time}.
As the figure shows, when processing a prompt with a length of around 15 tokens, \Sname takes less than 2.5 seconds to complete all three stages, which only constitutes a 14.7\% overhead to the total response time of ChatGPT.
The longest prompt we test has a length of around 1000 tokens, and \Sname takes less than 22.5 seconds to process it.
Because Llama 3's maximum input length is 4096 tokens, prompts longer than 3000 tokens (our topic query uses around 19 tokens) are truncated into multiple local LLM queries, increasing the processing time.

\subsubsection{CPU, Memory, and GPU Usage}

\begin{figure}
\centering
\includegraphics[width=0.4\textwidth]{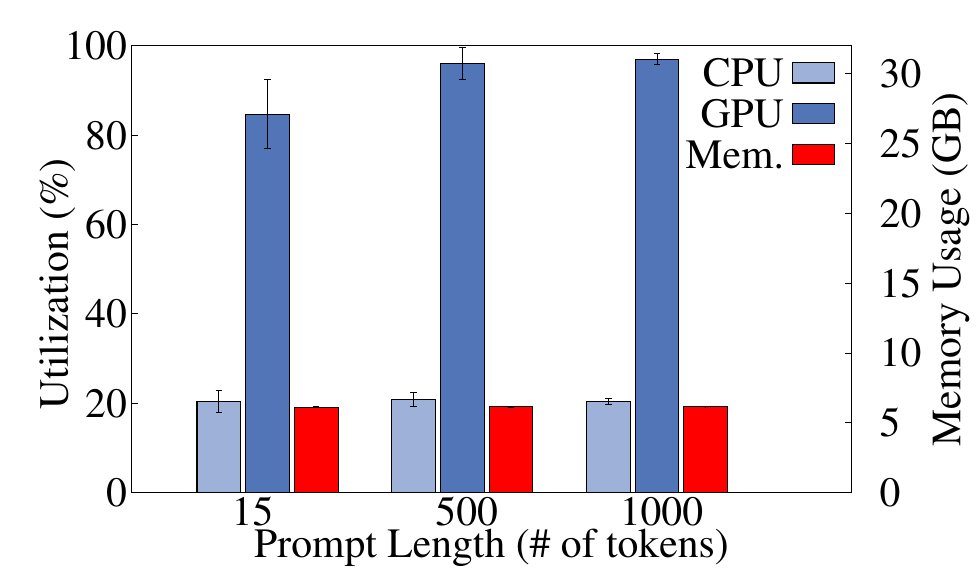}
\caption{CPU, GPU utilization and memory usage in processing prompts with various lengths in \Sname. Mem. stands for memory.}
\label{fig:processing_cpu}
\end{figure}

We measure the CPU, memory, and GPU usage of \Sname when processing the same set of synthesized prompts.
We utilize Activity Monitor for the measurements of all these various aspects by focusing on the Google Chrome Helper (GPU) process.
Both CPU and GPU usage are measured in percentage whereas memory usage is measured in megabytes.
For all these measurements, we record the peak consumption of each resource by the Google Chrome Helper (GPU) process.
The reason why we focus on this particular process is that the Google Chrome GPU process
runs both the ML-based named-entity model and local LLM inference.
The CPU usage is relatively low compared to the GPU usage, and the memory usage is stable throughout the processing of the prompts.
The reason for low CPU usage is that WebLLM is fully utilizing GPU for running the local LLM model.
On the other hand, the memory usage is accounted for by the size of the local LLM model that is loaded into the GPU memory.
Therefore, memory usage remains almost constant throughout the entire process at approximately 6.15GB.
The CPU and GPU utilization are reported in Figure~\ref{fig:processing_cpu}.

The results show that \Sname introduces a minimal CPU and memory overhead to user's device.
The CPU usage is around 20\% in a 10-core CPU machine.
Memory usage is mainly due to the size of the local LLM model that is loaded into the GPU memory.
The Llama 3 8B model we use in our experiment is around 6.15GB in size which provides high accuracy in sensitive topics detection.
Modern laptops are often equipped with 8GB - 36GB of integrated memory for CPU and GPU~\cite{macbook_spec}.
The laptop that we used in our experiment can handle \Sname without noticeable performance degradation.
Without \Sname, a user may load a local LLM model to process prompts locally to ensure privacy.
However, local LLM inference of a model at the size of OpenAI's ChatGPT model is prohibitively expensive to run on consumer-grade devices.

\section{Related Work}
\label{sec:related}

\subsection{Trusted and Confidential GPU Computing}

Trusted computing (also known as confidential computing) is an emerging technology that protects security-sensitive applications from privileged adversaries in the system stack.
CPU manufacturers, such as Intel and AMD, have introduced CPU features for running applications in secure enclaves, such as Intel SGX~\cite{mckeen2013} and AMD SEV~\cite{amdsev}.
SGXIO~\cite{weiser2017} extends Intel SGX to support I/O devices.
The Split-Trust hardware design~\cite{yao2023} performs physical isolation and static partition for I/O resources.
Graviton~\cite{volos2018} is a trusted computing framework that encrypts GPU workloads within an enclave and only decrypts them within the GPU.
ShadowNet~\cite{sun2023shadownet} leverages a CPU enclave to protect non-linear layers and only uses GPUs for linear layer computation.

As discussed in Section~\ref{sec:model},
browsers and \oses are trusted.
\Sname does not rely on trusted computing methods.
However, our system can benefit from trusted computing to further protect user's privacy and security.
For example, a privacy-sensitive prompt may use a local LLM running in a GPU enclave to perform topic identification.

\subsection{Browser GPU Stack Security}

Browser GPU stack is a
potential attack surface for malicious web apps to exploit the browser GPU stack and the kernel-mode GPU driver
vulnerabilities~\cite{peng2023}.
WebGL is a popular web API for rendering 2D and 3D graphics using JavaScript and it is supported by most modern browsers~\cite{yao2018}.
WebGPU is the next-generation API for GPU computing,
which generalizes WebGL and provides fine-grained control over GPU resources~\cite{webgpu}.
Previous research has shown that device drivers' code quality is poor: 85\% of the total Linux kernel bugs are in device drivers~\cite{vanderstoep2016} and they are difficult to find and fix~\cite{bursey2024syzretrospector, seyedtalebi2021}.
The convenience of WebGL and WebGPU introduces security risks, as they allow web apps to access privileged GPU code paths through a browser.
Several research works have studied WebGL graphics security.
Sugar~\cite{yao2018} leverages hardware-based GPU virtualization to protect \os from untrusted WebGL tasks.
Milkomeda~\cite{yao2018_2} studies the security checks from WebGL and repurposes them for mobile graphics security.

\Sname uses WebGPU to accelerate local LLM inference for topic identification.
We rely on the browser's security features to protect the confidentiality and integrity of local LLM inference.
For added security, \Sname can benefit from browser GPU security research,
but at the cost of increased complexity.

\subsection{Client-Side Filtering}

Removing sensitive information from client's inputs before sending them to a server is a common practice in client-server applications.
Client-side filtering has been used in voice assistants to prevent sensitive speech commands from being sent to third-party voice assistant skills~\cite{seyedtalebi2021_2} and in remote desktop applications to prevent sensitive information and control interface from being accessed by a remote party~\cite{liu2023methods}.

These solutions are orthogonal to \Sname, as they remove sensitive information before users' inputs are seen by a server.
However, \Sname focuses on the use case of online LLM services and addresses the unique challenges of filtering PIIs and privacy-sensitive topics in LLM prompts.

\subsection{Privacy-Preserving Machine Learning}

Privacy-preserving machine learning is a research area that aims to protect data privacy in machine learning training and inference.
Federated learning~\cite{mcmahan2017communication} is a distributed approach for training ML models on separate devices and only exchanges model updates instead of raw data that may be privacy-sensitive.
Homomorphic encryption~\cite{gentry2009fully} allows computation on encrypted data and it has been extended to support machine learning tasks~\cite{gilad2016cryptonets}.

As of today,
homomorphic encryption is not applicable to LLM inference and
federated learning LLM approaches~\cite{fan2023fate, zhang2024towards} require significant modifications to the LLM model and training process.
\Sname focuses on providing a lightweight solution in a client-server LLM inference setting.
\section{Discussion}
\label{sec:discussion}

\subsection{Users of Other Languages}

In our prototype of \Sname, we have focused on English-speaking users, but the system design applies to global users of cloud-based LLM services.
Keywords and patterns can be easily extended to other languages and the named-entity recognizer for other languages, such as Portuguese~\cite{souza2019portuguese} and Chinese~\cite{chang2021chinese}, are readily available.
Pre-trained local LLM models for other languages are also widely available online~\cite{hugging_face_lang}.
As discussed in Section \ref{sec:design_named_entity} and Section \ref{sec:design_llm_topic}, \Sname's architecture supports plug-and-play of named-entity recognition models and local LLM models.

\subsection{Performance Optimization}

Our prototype of \Sname demonstrates that the performance overhead of \Sname is minimal.
Several opportunities exist to further optimize the pattern matching stage, such as
using a rule-based modeling language for nonlinear pattern matching~\cite{warnke2021nonlinear}
or adopting a more efficient string-matching algorithm~\cite{aho1975efficient}.
For the named-entity detection stage, our architecture can readily support new named-entity recognition models, which may be more efficient.
For the local LLM inference stage, we expect the performance to improve as WebGPU matures (currently in an experimental phase~\cite{webgpu})
and as more efficient local LLM models are developed.
We also expect consumer-grade CPUs and GPUs to become more powerful, which will further
benefits the performance of \Sname.
\section{Conclusions}
\label{sec:conclusion}
We studied the privacy implications of
using online LLM services
and their third-party plugins.
To address this issue,
we presented \Sname,
a prompt sanitization technique
to protect user privacy by detecting and removing private and sensitive information from the users' prompts
before sending them to online LLM services.
\Sname is a browser extension designed to be a lightweight and efficient solution that can be easily integrated into existing systems
without any modifications to the web-based LLM services.
We demonstrated that \Sname can detect private and sensitive information from the users' prompts with high accuracy.
Furthermore, \Sname only incurs a small performance overhead which is suitable for real-world applications.
\section*{Acknowledgments}

This work has received no external funding.
The authors thank OpenAI for donating API credits to this project.

\balance
\bibliographystyle{unsrt}
\bibliography{zephyr}

\end{document}